\documentclass{epl}

\title{Thermodynamics of micellization of oppositely charged polymers}
\author{M. Castelnovo\inst{1}}
\institute{
  \inst{1} Department of Chemistry and Biochemistry, \\
University of California Los Angeles, Los Angeles, CA 90095}
\pacs{61.20.Qg}{Structure of associated liquids: electrolytes, molten salts, etc.}
\pacs{82.35.Rs}{Polyelectrolytes}
\pacs{87.15.Nn}{Properties of solutions; aggregation and crystallization of macromolecules}

\begin{document}

\maketitle

\begin{abstract}
The complexation of oppositely charged colloidal objects is considered in this paper as a thermodynamic micellization process where each kind of object needs the others to micellize. This requirement gives rise to quantitatively different behaviors than the so-called mixed-micellization where each specie can micellize separately. A simple model of the grand potential for micelles is proposed to corroborate the predictions of this general approach.
\end{abstract}

Complexation of oppositely charged colloidal objects has been the subject of continuous and important researches since the seventies because of the numerous applications ranging from flocculants and coatings to systems mimicking biological phenomena \cite{reviewcomplex}. The interest in this last field has grown in recent years as the drug carrier potential of polymer complexes has been realized \cite{kataokareview}: most of the non-viral vectors used for gene delivery are made of complexes between negatively charged DNA and oppositely charged objects that can be either surfactants (lipoplexes) \cite{safinya} or homo or co-polymers (polyplexes) \cite{behr,kataokareview}. Since a drug delivery vector has to be able to cross several biological and physical obstacles, the tailoring of a multifunction carrier challenges nowadays biologists as well as physico-chemists \cite{behr2}. The common feature of these complexes is their reliance on electrostatic self-assembly. Complementary to the experimental approaches looking for various drug delivery systems based on electrostatic complexation, there is obviously a need for more thorough understanding of the underlying mechanisms of self-assembly and of the relevant parameters for controlling them \cite{roland}.

In this Letter I present the general formalism required to describe the thermodynamic properties of micelles formed by complexation of oppositely charged objects and having a well defined gaussian size distribution. This allows us to address in particular the role of the mixture stoichiometry on the micelle compositions. In order to derive the most general features of the complexation viewed as a micellization process, the kind of objects considered will not be specified . The word ``micellization'' is used here in a generalized way: monodisperse amphiphilic molecules self-assemble forming micelles above the so-called critical micelle concentration as they are \textit{not soluble} as a single specie \cite{bill}. In the case of complexation, both components of the complexes are \textit{water-soluble}, but they still form aggregates that have been shown in the case of various polymer systems to exhibit a core-corona structure similar to the polymeric micelles made of neutral hydrophobic-hydrophilic copolymers \cite{kataokareview}. This extension of the micelle concept from aggregates of amphiphilic molecules to electrostatic complexes has been put forward quite recently by Cohen-Stuart \textit{et al.} \cite{cohen}. I will emphasize in this work the major differences between these two kinds of micellization \cite{pierre}. As an example, a phenomenological model of grand potential describing micelles made of oppositely charged objects is proposed and the phase behavior of such a solution is derived numerically. 

Depending on the kind of systems considered and on the stoichiometry of the mixture, two main features are observed \cite{reviewcomplex}: one can have either a macroscopic phase separation, associated with the neutralization of charges, or a partial micellization where each aggregate carries a net charge. For the sake of simplicity, I will focus mainly on the partial aggregation in this work . Physical realizations of such systems are, for example, complexes formed with polyelectrolytes and so-called double hydrophilic polymers -- diblock copolymers containing a polyelectrolyte block of opposite charge and a water-soluble neutral block (typically polyethylene glycol or PEG) \cite{kataokareview,kabanov,gohy,dautzenberg}. The macroscopic aggregation is limited by the neutral blocks forming a corona around a complexed core. This choice of systems allows us to focus on the effect of the stoichiometry of the solution on the properties of the micelles \cite{ralph}. Moreover, complexes of classical oppositely charged polyelectrolytes are shown to aggregate only partially for non neutral stoichiometries of the mixture and therefore some results obtained here might be also useful in this case. For strongly charged polyelectrolytes, the complexation is thought to take place out of thermodynamic equilibrium. Nevertheless, the knowledge of equilibrium properties is the first step towards kinetic or dynamical modeling.

Let us consider a bidisperse solution made of oppositely charged objects. Since I do not allow for macroscopic phase separation in the system, the small ions neutralizing the solution are not explicitly taken into account, to simplify the discussion; the effect of the salt can be thought to be incorporated into the free energy of the micelles. Each object is supposed to carry a $\pm 1$ charge. The energy unit is set to $kT=1$ for the remaining of the paper. The free energy density describing the solution is given in the dilute limit by
\begin{equation}
F=c_+ [\log c_+ +F_+]+c_- [\log c_- +F_-] +\sum_{p_+,p_-=1}^{\infty}c_{p_+,p_-}[\log c_{p_+,p_-}+F_{p_+,p_-}]\end{equation}
The quantities $c_+,c_-,c_{p_+,p_-}$ are respectively the concentrations of positive and negative unaggregated objects and the concentration of micelles made of $p_+$ and $p_-$ such objects. The corresponding excess chemical potentials are $F_+,F_-$ and $F_{p_+,p_-}$. This free energy also takes into account the translational entropy of single objects and micelles. The equilibrium state of the system is determined by its minimization with respect to all concentrations, and subject to the following mass conservation equations
\begin{eqnarray}
c_+ +\sum_{p_+,p_-=1}^{\infty}p_+c_{p_+,p_-}& =& \phi_+ \\
c_- +\sum_{p_+,p_-=1}^{\infty}p_-c_{p_+,p_-}& =& \phi_-
\end{eqnarray}
The total concentration of positive and negative objects are respectively $\phi_+$ and $\phi_-$.
Introducing the two Lagrange multipliers $\mu_+$ and $\mu_-$ corresponding to the mass conservation equations, the minimization equations can be written in the form:
\begin{eqnarray}
c_+ & = & \exp\{ -(F_+-\mu_+)\}\\
c_- & = & \exp\{ -(F_--\mu_-)\}\\
c_{p_+,p_-} & = & \exp\{-(F_{p_+,p_-}-\mu_+p_+ - \mu_-p_-)\}
\end{eqnarray}
$\mu_+$ and $\mu_-$ are interpreted as the chemical potential of positive and negative objects. 
The last equation shows how the concentration of micelles is related to the grand potential $\Omega_{p_+,p_-}=F_{p_+,p_-}-p_+\mu_+-p_-\mu_-$. 

In the case studied here of well defined micelles having a gaussian size distribution, the grand potential has a minimum located at the finite values $\bar{p}_+$ and $\bar{p}_-$ above some critial value for the chemical potentials. Therefore the sums in the mass conservation equations can be approximated by a steepest descent method. This amounts to considering an effective two state model for the components in the solution, the unaggregated object and the micelle state. The mass conservation equations can then be rewritten
\begin{eqnarray}
\label{mce+}1+\exp\{-\Omega_{\bar{p}_+,\bar{p}_-}\}\exp\{F_+-\mu_+\}\bar{p}_+ \Delta& = & \exp\{-(\mu_+ -\mu_+ ^0)\}\\
\label{mce-}1+\exp\{-\Omega_{\bar{p}_+,\bar{p}_-}\}\exp\{F_--\mu_-\}\bar{p}_- \Delta& = & \exp\{-(\mu_- -\mu_- ^0)\}
\end{eqnarray}
with the following notations $\Delta=2\pi/\sqrt{\delta_+\delta_--\delta_{\pm}^2}$, $\delta_i\stackrel{_{i=\pm}}{=}(\partial^2 F_{p_+,p_-}/\partial p_i^2)_{\bar{p}_+,\bar{p}_-}$ and $\delta_{\pm}=(\partial^2 F_{p_+,p_-}/\partial p_+\partial p_-)$ taking into account the width of the micelle size distribution. Each total concentration has been rewritten in the preceding equations by introducing the chemical potential of objects without any association $\mu_i^0=\log\phi_i+F_i$. 

The solutions of Eq.\ref{mce+} and \ref{mce-} for $\mu_+$ and $\mu_-$ together with the minimization equations for the grand potential allows us to determine all the equilibrium properties of the solution, \textit{e.g.} to draw its phase diagram, once the grand potential of the micelles is known. One can understand the micellization behavior in the solution by following particular paths in the phase diagram. This is done numerically below with a crude model of grand potential bearing the two essential features of the particular micellization studied here: one requires that \textit{(i)} micelles cannot form if only one component is present in solution and that \textit{(ii)} any net charge of the micelles has an energetic price. The first requirement can be easily taken into account by an attractive term of the form $F_{att}=Ap_+^{\alpha}p_-^{\beta}$. For positive values of $A$, the sum of the exponents has to be lower than 1, so that this form describes an attraction. This can be thought as a ``surface tension'' term favoring the growth of the complex, similar to the micelles made of regular diblock hydrophilic/hydrophobic copolymers, although strictly speaking such a term would be proportional to the two-third power of the sum of aggregation numbers ( which is unable to fulfill requirement \textit{(i)}). Using nevertheless this analogy, the exponents are chosen to be $\alpha=\beta=1/3$ for a symmetric system, so that this term favoring aggregation corresponds exactly to the single micellization case in the case of zero net charge. Note that the precise values of the exponents do not change the qualitative behavior described below. The second requirement can be taken into account with a quadratic energetic price for net charges, $F_{asym}=B(p_+-p_-)^2$. Such a form occurs for example simply by taking into account the loss of translational entropy by small ions neutralizing the excess charge in the complex \cite{ralph,casteldpa}. Finally, the prohibition of macroscopic phase separation through the complexation is ensured by a term representing for example the energetic cost of a polymeric corona around the complexed core, $F_{corona}=C(p_+ +p_-)^{3/2}$ \cite{daoud}. The choice of a corona-like stabilization of the complex is obvious for cases where there is actually a corona, while it might look odd for a mixture of oppositely charged homopolyelectrolytes. Nevertheless, as proposed in reference \cite{casteldpa} for diblock polyampholytes, the excess charge might be favorably located on arms surrounding the neutral complexed core. 

\begin{figure}
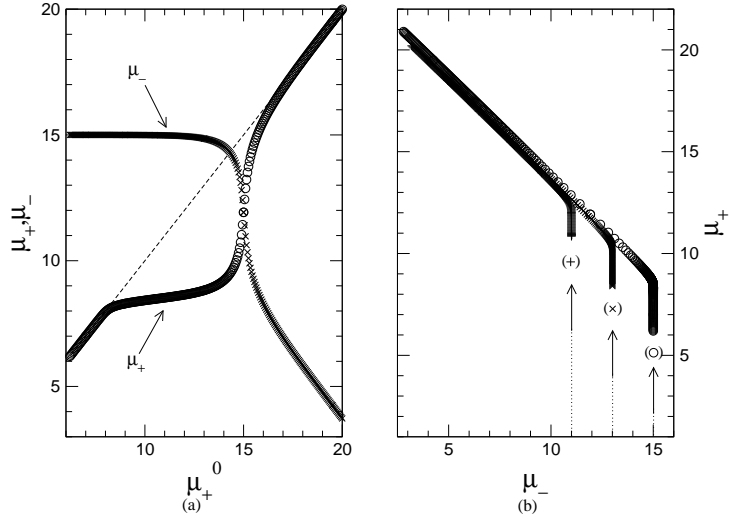

\onefigure{figure1.eps}
\caption{(a) Chemical potential of positive (circles) and negative (crosses) objects as a function of the bare chemical potential 
of positive objects $\mu_+^0$ for $\mu_-^0=15$. The dashed line represents $\mu_+=\mu_+^0$. (b) Particular paths in the phase diagram for different starting points (also indicated by arrows): $\mu_-^0=11$ (pluses), $\mu_-^0=13$ (crosses), $\mu_-^0=15$ (circles). The parameters of the grand potential have been set equal to $A=30,\, B=10,\, C=1,\, F_+=F_-=0$.}
\label{fig1}
\end{figure}

The grand potential is therefore written as
\begin{equation}
\label{grand}\Omega_{p_+,p_-}=Ap_+^{1/3}p_-^{1/3}+B(p_+-p_-)^2+C(p_+ +p_-)^{3/2}-\mu_+p_+-\mu_-p_-
\end{equation}
The mass conservation equations Eq.\ref{mce+} and \ref{mce-} are then solved numerically for a given set of parameters together with the minimization equations of the grand potential to obtain the chemical potentials of positive and negative objects. As will be shown below, the results can be traced back to the general properties of this specific kind of micellization. The properties of the solution can be probed through particular paths in the phase diagram. Experimentally relevant paths correspond for example to the increase of the number of positive objects at fixed number of negative objects. An example of numerical solution for the chemical potentials along such a path is given on figure \ref{fig1}.
At very low concentration of positive objects, there is no association and $\mu_{\pm}=\mu_{\pm}^0$. Above
a critical concentration of positive objects, micelles start to form and the chemical potential of positive objects saturates while the chemical potential of negative objects remains unaffected: almost every added positive object is consumed by forming a complex with negative objects. This situation is very similar to the monodiperse micellization case of amphiphilic molecules, where the chemical potential of molecules is almost constant above the critical micellar concentration (CMC). The negative objects play in the case of the present study the role of ``linkers'' that help the positive objects to aggregate. The nearly constant value of chemical potential of positive objects, which gives also the onset of micellization, can be found by requiring that the term associated with the micelles in Eq.\ref{mce+} is of order one:
\begin{equation}
\mu_+ \simeq\frac{F_{\bar{p}_+,\bar{p}_-}-F_+}{\bar{p}_+-1}-\frac{\log \left[\bar{p}_+ \Delta\right]}{\bar{p}_+-1}-\frac{\mu^0_- \bar{p}_-}{\bar{p}_+-1}
\end{equation}
Just above this onset of micellization, most of the negative objects are still in the unaggregated state. The micelle term in the mass conservation equation of negative objects becomes of order of the isolated object term for a higher value of $\mu_+^0$. Using Eq.\ref{mce+}, the micelle term in the mass conservation equation of negative objects is rewritten just above the onset of micellization as
\begin{equation}
\exp\{-\Omega_{\bar{p}_+,\bar{p}_-}\}\exp\{F_--\mu_-\}\bar{p}_- \Delta=\frac{\phi_+\bar{p}_-}{\phi_-\bar{p}_+}\end{equation}

Since the micelles under consideration are such that they cannot handle large non-neutral stoichiometric ratio $\bar{p}_+/\bar{p}_-\neq 1$, the micelles start to dominate the behavior of negative objects very close to the value $\mu_+^0=\mu_-^0 -F_-+F_+$, which corresponds to the equality of concentrations of positive and negative objects. This particular value is due to the unit charge carried by each object. The straightforward generalization of this result to objects of charge $q_+$ and $q_-$ is given by $\mu_+^0=\mu_-^0 -F_-+F_++\log(q_+/q_-)$. Above this second critical value, the consumption of added positive objects in complexes can not be as high as in the regime of excess of negative objects. Therefore the unaggregated objects dominate once again the behavior of positive objects, and their chemical potential is asymptotically given by $\mu_+=\mu_+^0$. The chemical potential of negative objects reaches the asymptotic value 
\begin{equation}
\mu_-=\frac{F_{\bar{p}_+,\bar{p}_-}-F_-}{\bar{p}_-}-\frac{\log \bar{p}_- \Delta}{\bar{p}_-}-\frac{\bar{p}_+}{\bar{p}_-}\mu_+^0
\end{equation}
The decrease in the chemical potential of negative objects is associated with the energetic gain of complex formation and the increasing number of negative objects involved. All these features are indeed observed for the proposed grand potential through the numerical solutions. Different starting concentrations of negative objects are shown in figure \ref{fig1} to end up roughly on the same ``micellization'' line of the phase diagram ($\mu_+,\mu_-$), which corresponds approximately to a constant value of the grand potential minimum 
\begin{equation}
\Omega_{\bar{p}_+,\bar{p}_-}\simeq F_-+\log\bar{p}_-\Delta-\mu_-^0
\end{equation}

\begin{figure}
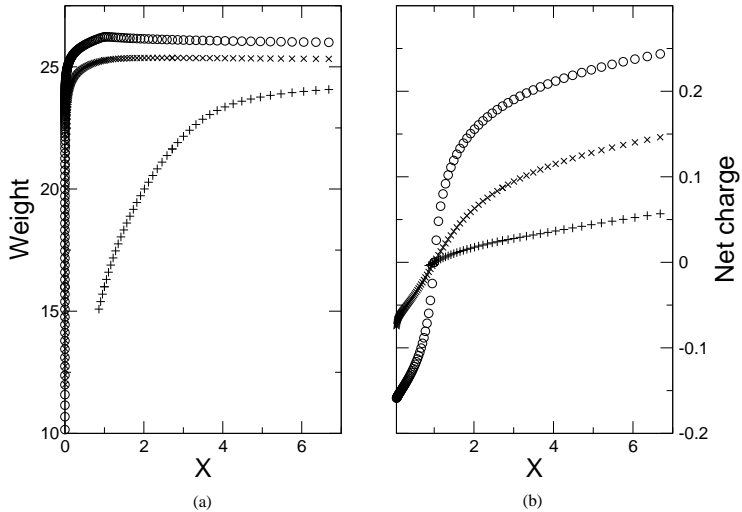

\onefigure{figure2.eps}
\caption{Structural parameters of the micelles as a function of the mixing ratio $X=\phi_+/\phi_-$ for different starting points. (a) Weight $\bar{p}_++\bar{p}_-$ and (b) net charge of the micelles $\bar{p}_+-\bar{p}_-$. Starting points: $\mu_-^0=11$ (pluses), $\mu_-^0=13$ (crosses), $\mu_-^0=15$ (circles). The parameters of the grand potential have been set equal to $A=30,\, B=10,\, C=1,\, F_+=F_-=0.$}
\label{fig2}
\end{figure}

In the case of bidipserse micellization where the two components can micellize separately, the behavior is quantitatively different for the same type of path in the phase diagram \cite{pierre}: above the critical micellar concentration the stoichiometry of the micelles follows the stoichiometry of the mixture, since asymmetry in the micelle composition costs essentially nothing in this case. As the concentration of one of the components is increased, the composition of the micelles gets close to the one of pure micelles of this component, and both chemical potentials saturate. Further increase of the concentration leads to an increase in the number of pure micelles of the added component. Moreover it has been shown that under certain conditions, the mixed micellization occurs in a discontinuous way, the aggregation number of one component jumping from 0 to a finite value $p_i \neq 1$ in already preformed micelles of the other components \cite{pierre}. This leads to a bimodal distribution of micelles. Such a phenomenon is not predicted in the case of the micellization of oppositely charged polymers.

Of more interest from the standpoint of the drug carrier potential of complexes are quantities like the net charge and the weight of the complexes. These are shown for the grand potential Eq.\ref{grand} as a function of the mixing ratio $X=\phi_+/\phi_-$ in figure \ref{fig2}. A unit mass for each object has been chosen for the sake of simplicity. The representation as a function of $X$ is often used to plot experimental data on complexes. As shown on figure \ref{fig2}, the net charge and the weight of the complexes depend not only on the mixing ratio, but also on the overall concentration, a fact that is indeed observed experimentally. The largest weight is achieved for neutral stoichiometry and increasing overall concentration \cite{dautzenberg}. Note that the saturation in the weight of the micelle, if any, is quite slow as a function of the overall concentration of the solution. The net charge of the micelles is shown to be tuned for the symmetric system studied here by the mixing ratio of the solution, although the effect is small here for the range of chosen parameters. For $\mu_-^0 =11$, the system is very close to the minimal concentration of negative objects required to grow any complexes; this explains the slow increase of both the mass and the net charge. These results still show that one must be careful in modeling overcharging phenomena in complexes: studies like the one of Kunze and Netz on complexes formed by one DNA with excess flexible polycations can be naturally complemented by the formalism used here to address for example the effect of finite DNA concentrations \cite{roland}.

The present study provides a rigorous framework to study the thermodynamic behavior of micelles made of oppositely charged polymers and in particular the effect of concentration on the main features of micelles. The crude choice of grand potential, based on the requirement that it describes in a phenomenological way species that need each other to micellize, exhibits the qualitative features of polyelectrolyte complexation \cite{dautzenberg}. Nevertheless, a more precise choice of grand potential, as well as a more thorough investigation of the relation between effective, \textit{i.e.} $A,B,C$, and physical parameters (salt amount, charge fraction, etc...) is needed to address for structural aspects ( \textit{e.g.} radii) of micelles as well as the effect of salt on the general properties of micelles. This is the topic of further works in progress.

\acknowledgments I would like to thank P. Sens (ICS, Strasbourg) for fruitful discussions about mixed micellizations and J.-F. Joanny (Institut Curie, Paris) and W. M. Gelbart (UCLA) for critical reading of the manuscript. This work was partially supported by NSF grant \# CHE9988651 to W.M.G.

\end{document}